\documentclass{ifacconf}

\usepackage{graphicx}      
\usepackage{natbib}        
\usepackage{amsmath}
\usepackage{amsfonts}

\begin{document}
\begin{frontmatter}

\title{Computational Efficiency of Fractional Diffusion Using Adaptive Time Step Memory\thanksref{footnoteinfo}} 

\thanks[footnoteinfo]{This work was funded by NIH grant RO1 NS054736. C.M. was also partially supported through NIH T32 EB009380. }

\author[First]{Brian P. Sprouse}$^\dagger$
\author[Second]{Christopher L. MacDonald}$^\dagger$
\author[Third]{Gabriel A. Silva}

{\it  Department of Bioengineering,\\University of California, San Diego, La Jolla, CA 92093 USA}
\address[First]{(e-mail: bsprouse@ucsd.edu)}
\address[Second]{(e-mail: clmacdon@ucsd.edu)}
\address[Third]{(e-mail: gsilva@ucsd.edu)}

{\it $^\dagger$ These authors contributed equally to this work}

\begin{abstract}                
Numerical solutions to fractional differential equations can be extremely computationally intensive due to the effect of non-local derivatives in which all previous time points contribute to the current iteration. In finite difference methods this has been approximated using the 'short memory effect' where it is assumed that previous events prior to some certain time point are insignificant and thus not calculated. Here we present an adaptive time method for smooth functions that is computationally efficient and results in smaller errors during numerical simulations. Sampled points along the system's history at progressively longer intervals are assumed to reflect the values of neighboring time points. By including progressively fewer points as a function of time, a temporally `weighted' history is computed that includes contributions from the entire past of the system, resulting in increased accuracy, but with fewer points actually calculated, which ensures computational efficiency.
\end{abstract}


\end{frontmatter}

\section{Introduction}
\label{Introduction}
Throughout the past few decades, fractional differential equations have played an increasing role in the disciplines of physics (\cite{PhysRevE.55.6821,fellah2002application,metzler1999anomalous,soczkffiwicz2002application}), chemistry (\cite{gorenflo2002time,seki2003fractional}), and other engineering and science fields (\cite{cushman36fractional,baeumer2001subordinated,mathieu2003fractional}). In particular, fractional calculus methods offer unique approaches for modeling complex and dynamic processes at the cellular level. Recent research using these methods have been applied to modeling ultrasonic wave propagation in cancellous bone (\cite{sebaa2006application}), bio-electrode behavior at the cardiac tissue interface (\cite{magin2008modeling}), and describing spiny dendrite neurons through a fractional cable equation (\cite{henry2008fractional}). Other work has shown that endogenous lipid granules within yeast exhibit anomalous subdiffusion during mitotic cell divison (\cite{selhuber2009variety}).
 

Here we investigate the implementation and subsequent computational performance of a fractional reaction-diffusion equation (FRD) in numerical simulations. We introduce an `adaptive time step' method for computing the contribution of the memory effect associated with the history of a system, which fractional methods must take into account.  While this method can be applied to any diffusion application modeled as a fractional process, it offers particular significance for modeling complex processes with long histories, such as lengthy molecular or cellular simulations. Approaches for numerically approximating solutions to fractional diffusion equations have been extensively studied (\cite{chen2008finite,langlands2005accuracy,liu2007stability,mainardi2006sub,mclean2009convergence,podlubny2009matrix,yuste2003explicit}), but in general there is always a trade off between computational efficiency and the accuracy of the resultant approximations. We developed a simple approach that significantly reduces simulation times but which does not significantly affect the computed accuracy of the final solution. This was accomplished through an `adaptive time step memory' method by changing the interval of the backwards time summation in the Gr\"{u}nwald-Letnikov fractional derivative formulation used in many numerical schemes for discretizing fractional diffusion equations from a continuous Riemann-Liouville approach (\cite{chen2009numerical,yuste2003explicit,podlubny1999fractional}). Instead of incorporating every previous time point into the most current calculation, only certain points were recalled based upon their proximity to the current point in time and weighted according to the sparsity of the time points. This substantially reduced the computational overhead needed to recalculate the prior history at every time step, yet maintained a high level of accuracy compared to other approximation methods. We compared this new `adaptive time' approach with the `short memory' principle described by Volterra (\cite{volterra1931}) and Podlubny et al (\cite{podlubny1999fractional}) and examined differences in simulation times and errors under similar conditions for both methods. 

All calculations and simulations in the results were based on an explicit numerical method utilizing the Gr\"{u}nwald-Letnikov relationship for the fractional reaction diffusion equation given by:

\begin{equation}
\label{fracdiff}
\frac{d^{\gamma}u(\vec x,t)}{dt^\gamma}=\alpha\nabla^2u(\vec x,t)+f(u)~~~~\vec x=\left\{x_1,x_2...x_N\right\}
\end{equation}

where $\gamma$ is the anomalous diffusion exponent, $\vec x$ is an N dimensional vector, $\alpha$ is the coefficient of diffusivity, and \emph{f(u)} represents a consumption or generation term.

\section{Methods}

\subsection{Gr\"{u}nwald-Letnikov formulation and simplification}

Consider the fractional diffusion equation with no reaction term,

\begin{equation}
\frac{d^{\gamma}u}{dt^\gamma}=\alpha\nabla^2u
\end{equation}

where $\alpha$ is the diffusion coefficient. Using the relation

\begin{equation}
\frac{du}{dt}=\frac{d^{1-\gamma}u}{dt^{1-\gamma}}\frac{d^{\gamma}u}{dt^\gamma}=>\frac{du}{dt}=\frac{d^{1-\gamma}u}{dt^{1-\gamma}}\alpha\nabla^2u
\end{equation}

the diffusion equation then becomes

\begin{equation}
\frac{du}{dt}=\alpha D^{1-\gamma}\nabla^2u
\label{diffusioneq}
\end{equation}

where $D^{1-\gamma}$ represents the continuous time Riemann-Liouville fractional derivative. We then implement the Gr\"{u}nwald-Letnikov definition for numerical simulations.  This approach discretizes the Riemann-Liouville derivative through the relationship:

\begin{equation}
D^{n}u=\lim_{\tau\rightarrow0}\tau^{n}\sum_{m=0}^{t/\tau}(-1)^m {n\choose m}f(t-m\tau)
\label{grunletdef}
\end{equation}

where ${n\choose m}$ represents the binomial coefficient given by

\begin{equation}
\label{binom}
 {n\choose m}=\frac{n!}{m!(n-m)!}=\frac{\Gamma(n+1)}{m!\Gamma(n+1-m)}.
\end{equation}

Applying eq. \ref{grunletdef} to the fractional diffusion equation (eq. \ref{diffusioneq}) yields

\begin{equation}
 \frac{du}{dt}=\lim_{\tau\rightarrow0}\tau^{\gamma-1}\alpha\sum_{m=0}^{t/\tau}(-1)^m {1-\gamma\choose m} \nabla^2u(t-m\tau).
\label{grunlet_equation}
\end{equation}

From here, one can simplify the binomial coefficient to a simple recursive relation. 

Next, define a function $\psi(\gamma,m)$ such that

\begin{equation}
\label{psidefinition}
 \psi(\gamma,m)=(-1)^m {1-\gamma\choose m}
\end{equation}

and substituting ${1-\gamma}\choose m$ into eq. \ref{binom} yields

\begin{equation}
\psi(\gamma,m)=\frac{(-1^m)\Gamma(2-\gamma)}{m!\Gamma(2-\gamma-m)}.
\end{equation}

Substituting the relation $\Gamma(n)=(n-1)\Gamma(n-1)$ into eq. \ref{psidefinition} results in

\begin{equation}
 \psi(\gamma,m)=\frac{(-1)^{m-1}\Gamma(2-\gamma)}{(m-1)!\Gamma(2-\gamma-(m-1))}\frac{-1}{m\frac{1}{2-\gamma-m}}
\end{equation}

yielding the iterative relationship

\begin{equation}
 \psi(\gamma,m)=-\psi(\gamma,m-1)\frac{2-\gamma-m}{m}.
\end{equation}

For $m = 0$,

\begin{equation}
 \psi(\gamma,0)=\frac{(-1^0)\Gamma(2-\gamma)}{0!\Gamma(2-\gamma-0)}=1.
\end{equation}

This recursive function is valid for all $\gamma$ including subdiffusion, standard diffusion and superdiffusion, so this equation is general over all regimes. Since there is no use of $\Gamma(x)$, there are no values which tend to infinity or limits to compute. In addition, since these values are used over and over again during the course of a numerical simulation, they can be precomputed for values of $m=0$ to $m=N$, where $N$ is the number of time points, resulting in a significant savings in performance. A similar simplification has been previously discussed and used in \cite{podlubny1999fractional}.

\subsection{Discretization}

The substitution of $\psi(\gamma,m)$ into eq. \ref{grunlet_equation} results in

\begin{equation}
\frac{du}{dt}=\lim_{\tau\rightarrow0}\tau^{\gamma-1}\alpha\sum_{m=0}^{t/\tau}\psi(\gamma,m) \nabla^2u(t-m\tau).
\end{equation}

Then, using 

\begin{equation}
 \nabla^2u(t-m\tau)=\frac{\partial^2u(\vec x,t-m\tau)}{\partial x_1^2}+\frac{\partial^2u(\vec x,t-m\tau)}{\partial x_2^2}
\end{equation}

as the formulation in two dimensions, one can discretize the function into a finite difference based FTCS scheme (forward time centered space) on a grid $u_{j,l}^k$ where $k=t/\Delta_t,j=x_1/\Delta_{x},l=x_2/\Delta_{x}$, using the relations (\cite{press1992numerical})

\begin{eqnarray}
\frac{\partial^2u(\vec x,t-m\tau)}{\partial x_1^2}&=&\frac{u_{j+1,l}^{k-m}-2u_{j,l}^{k-m}+u_{j-1,l}^{k-m}}{\Delta_{x}^2}\nonumber \\
\frac{\partial^2u(\vec x,t-m\tau)}{\partial x_2^2}&=&\frac{u_{j,l+1}^{k-m}-2u_{j,l}^{k-m}+u_{j,l-1}^{k-m}}{\Delta_{x}^2}\nonumber \\
\frac{\partial u(\vec x,t)}{\partial t}&=&\frac{u^{k+1}_{j,l}-u^{k}_{j,l}}{\Delta_t}.
\end{eqnarray}

In the discrete limit where $\tau\rightarrow\Delta_t$ 

\begin{equation}
\label{frac_norxn}
\frac{u^{k+1}_{j,l}-u^{k}_{j,l}}{\Delta_t}=\alpha\frac{\Delta_t^{\gamma-1}}{\Delta_{x}^2}\sum_{m=0}^k\psi(\gamma,m)\delta^{k-m}_{j,l}
\end{equation}

where $\delta^{k-m}_{j,l}$ is the finite difference kernel given by

\begin{equation}
 \label{deltasimp}
\delta^{k-m}_{j,l}=\left(u_{j+1,l}^{k-m}+u_{j-1,l}^{k-m}-4u_{j,l}^{k-m}+u_{j,l+1}^{k-m}+u_{j,l-1}^{k-m}\right).
\end{equation}

Adding a consumption/generation term is straightforward in this implementation. For example, take an exponential decay term given by

\begin{equation}
\label{expdecay}
 \frac{du}{dt}=-\beta u
\end{equation}

with the complementary finite difference relation

\begin{equation}\label{eq:findif}
 \frac{u^{k+1}_{j,l}-u^{k}_{j,l}}{\Delta_t}=-\beta u_{j,l}^{k}\Delta_t.
\end{equation}

Incorporating eq. \ref{expdecay} into the form of eq. \ref{diffusioneq} results in 

\begin{equation}
 \frac{du}{dt}=\frac{d^{\gamma-1}}{dt^{\gamma-1}}\nabla^2u-\beta u,
\end{equation}

which gives the full finite difference implementation in two dimensions

\begin{equation}
\label{frac_final}
\frac{u^{k+1}_{j,l}-u^{k}_{j,l}}{\Delta_t}=-\beta u_{j,l}^{k}+\alpha\frac{\Delta_t^{\gamma-1}}{\Delta_x^2}\sum_{m=0}^k\psi(\gamma,m)\delta^{k-m}_{j,l}.
\end{equation}

\section{Results}

\subsection{Diffusion simulation examples}

\begin{figure}
\begin{center}
\includegraphics[width=8.8cm]{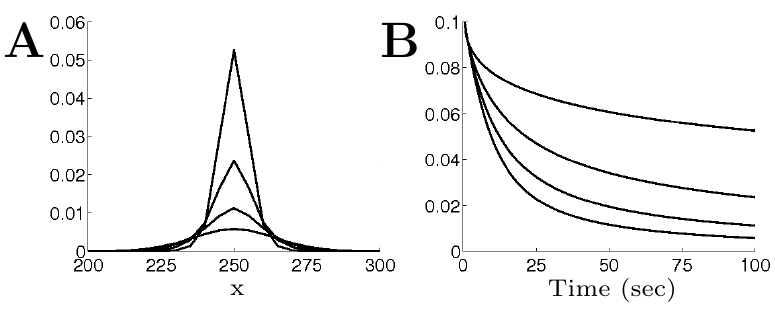}    
\caption{Simulation results for $\gamma=0.5,~0.75,~0.9,~1.0$ (for traces from top to bottom) in one dimensional space (panel A) and time (panel B). As the subdiffusion regime is approached the profile becomes more and more hypergaussian.} 
\label{diffresults}
\end{center}
\end{figure}

Fig.~\ref{diffresults} shows the results of simulating the equations for four different values of $\gamma$. Simluations were run on a 100 x 100 grid with initial conditions $U_{50,50}^0$ = 0.1, $U_{51,50}^0=U_{50,51}^0=U_{49,50}^0=U_{50,49}^0$ = 0.05 and zero elsewhere. Dirichlet boundary conditions were implemented and the simulation edge set to zero. In these simulations, $\alpha=1$, $\beta=0$, $\Delta_t=0.5$, $\Delta_x=5$, and run until $t=100$.

\subsection{Adaptive Memory Method}

\begin{figure}
\begin{center}
\includegraphics[width=8.8cm]{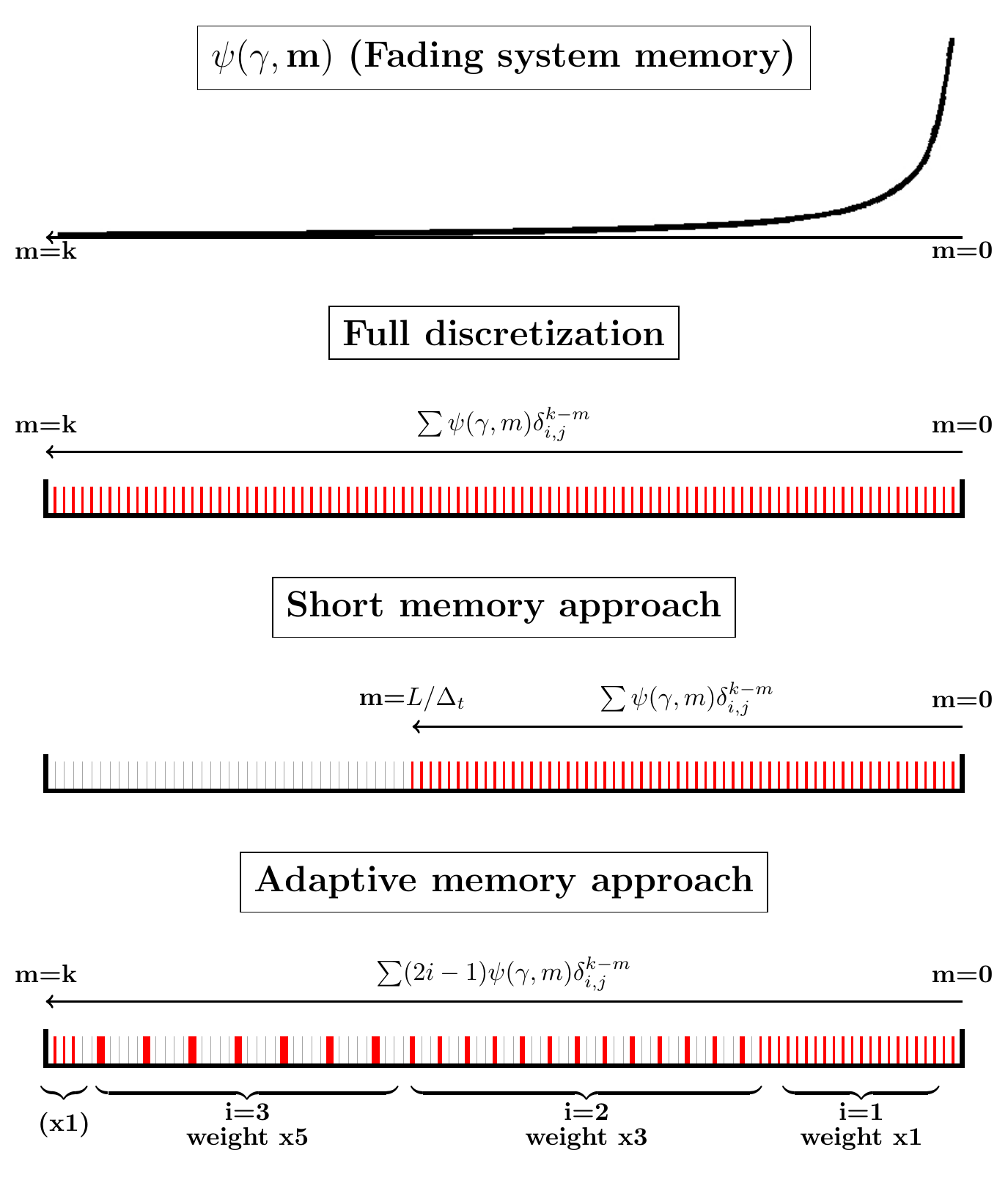}    
\caption{Short memory and adaptive memory methods for estimating the the Gr\"{u}nwald-Letnikov discretization. Both approximations rely on the sharply decreasing function $\psi(\gamma,m)$ as $m$ increases to omit points from the backwards summation. While short memory defines a sharp cut off of points, adaptive memory provides a weighted sampling of points for the entire history of the system. Points included in the computation by each method are highlighted in red. See text for details.} 
\label{adaptivetime}
\end{center}
\end{figure}

In eq. \ref{frac_final} each iteration requires the re-calculation and summation of every previous time point convolved with $\psi(\gamma,m)$.  This becomes increasingly cumbersome for large times, which require significant numbers of computations and memory storage requirements. To address this, Podlubny et. al (\cite{podlubny1999fractional}) introduced the `short memory' principle which assumes that for a large $t$ the role of the previous steps or `history' of the equation become less and less important as the convolved $\psi(\gamma,m)$ shrinks towards zero. This would then result in approximating eq. \ref{frac_final} by truncating the backwards summation, only taking into account times on the interval $[t-L,t]$ instead of $[0,t]$, where $L$ is defined as the `memory length' (eq. 7.4 in \cite{podlubny1999fractional}; Fig \ref{adaptivetime}). While computationally efficient, this approach leads to errors in the final solution since not all points are counted in the summation. Despite the resultant errors, this numerical method represents a powerful and simple approach for providing a reasonable trade off between computational overhead and numerical accuracy. In the context of the implementation derived here, it would result in a discretization scheme given by 

\begin{equation}
\label{frac_shortmemory}
\frac{u^{k+1}_{j,l}-u^{k}_{j,l}}{\Delta_t}=-\beta u_{j,l}^{k}+\alpha\frac{\Delta_t^{\gamma-1}}{\Delta_x^2}\sum_{m=0}^{min(L/\Delta_t,k)}\psi(\gamma,m)\delta^{k-m}_{j,l}.
\end{equation}

As an alternative to the method of Podlubny, we propose an adaptive time approach that also shortens computational times but which at the same time results in much greater accuracy than the use of `short memory'.  We achieve this by introducing the concept of an `adaptive memory' into the Gr\"{u}nwald-Letnikov discretization.  

The underlying principle of the adaptive time approach is that relative to the current time point previous time points contribute different amounts to the summation. Values relatively closer to the current time point will have a greater contribution to the current numerical calculation than values many time points back due to the multiplier $\psi(\gamma,m)$.  For smooth functions, as $m$ increases and $\psi(\gamma,m)$ decreases, neighboring points in the summation exhibit only small differences. Consequently, one can take advantage of this and utilize an `adaptive memory' approach in which neighboring values at prior time points are grouped together for defined increments and the median taken as a representative contribution for the increment and weighted according to the increment length to account for the skipped points. This results in fewer time points and fewer computations in the summation. Algorithmically, for an arbitrary $a$ time steps back from the current time point $k$ for which the history of the system is being computed, consider an initial interval $[0,a]$ for which all time points within this interval are used in the summation and therefore contribute to the the Gr\"{u}nwald-Letnikov discretization. Let subsequent time intervals become longer the further away from $k$ they are and within them only every other $d$ time points are used in the summation term, i.e. only the median values of a temporally shifting calculation increment of length $d$ along the current interval are considered. As subsequent intervals become longer so does $d$, thereby requiring less points to compute. 

\textbf{Definition 3.1} Let $k$ be the current iterative time step in a reaction diffusion process for which a Gr\"{u}nwald-Letnikov discretization is being computed. Consider an arbitrary time point $a$ in the history of the system backwards from $k$. For $i\in \mathbb{N}_1,i\neq 1$, define an interval of this past history by
\begin{equation}\label{eq:interval}
I=[a^{i-1}+i,a^i]
\end{equation}
where $\mathbb{N}_1$ represents the set of natural numbers beginning from one.  Given how the indices $i$ are defined, the very first interval backwards from $k$ is independent of \ref{eq:interval} and is given by $[0,a]$. This is considered the base interval. Subsequent intervals are defined as a function of this base, i.e. as a function of $a$ and determined by \ref{eq:interval}. 

Let $i_{max}$ be the value of $i$ such that $k \in I_{max}=[a^{i_{max}-1}+i_{max},a^{i_{max}}]$. The complete set of intervals then is defined as $\zeta=\{I=[a^{i-1}+i,a^i]:i\in \mathbb{N}_1,i\neq 1, i\le i_{max}\}$. 

\textbf{Definition 3.2} For the set of intervals $\zeta$ as defined in Definition 3.1, $D=\{d=2i-1:i\in \mathbb{N}_1,i\neq 1\}$ is the set of distances $d$ by which the corresponding intervals in $\zeta$ are sampled at.

\textbf{Proposition 3.1} In general, for two dimensional diffusion without  consumption or generation terms for any interval as defined in Definition 3.1, the Gr\"{u}nwald-Letnikov discretization with adaptive time step memory is given by 
\begin{align}\label{eq:adapttime}
& \frac{u^{k+1}_{j,l}-u^{k}_{j,l}}{\Delta_t}=\alpha\frac{\Delta_t^{\gamma-1}}{\Delta_x^2}\Bigg[\sum_{n=0}^{a}\psi(\gamma,n)\delta^{k-n}_{j,l}+\cdots \\ \nonumber
&\sum_{i=2}^{i_{max}}\sum_{m_i=a^{i-1}+i}^{a^i}(2i-1)\psi(\gamma,m_i)\delta^{k-m_i}_{j,l}+\sum_{p=m_{max}+i}^{k}\psi(\gamma,p)\delta^{k-p}_{j,l}\Bigg]
\end{align}
where $p \in \mathbb{N}_1$, and for each $i$ (i.e. for each interval) $M=\{m_i=a^{i-1}+(2i-1)\eta-i+1:\eta\in\mathbb{N}_1~\&~m_i\le m_{max}\}$ is the set of time points over which $\psi(\gamma,m)\delta^{k-m}_{j,l}$ is evaluated. Since the time point $k$ may be less than the full length of the last interval $I_{max}$, $|m_{max}| \le |m_{k-i}|$ represents the maximum value in $I_{max}$ that is evaluated, i.e. the last element in the set $M$ for $I_{max}$.

\textbf{Proof} The first summation represents the basis interval and the limits of the summation operator imply the contribution of every time point, i.e. $n\in\mathbb{N}_1$. For intervals beyond $a$: any arithmetic sequence defined by a recursive process $\nu_\eta=\nu_{\eta-1} + d$, $\eta\in\mathbb{N}_1$ for some distance $d$, the $\eta^{th}$ value in the sequence can be explicitly calculated as $\nu_\eta=\nu_1+(\eta-1)d$ given knowledge of the sequence's starting value $\nu_1$. For the set $\zeta$ this implies that $\nu_1=a^{i-1}+i$ and $d=2i-1$ for a given $i$. This then yields $\nu_\eta=a^{i-1}+i+(\eta-1)(2i-1)=a^{i-1}+(2i-1)\eta-i+1\mathrel{\mathop:}= m_i$ as required. The outer summation collects the summations of all intervals that belong to the set $\zeta$.  The last summation on the interval $[m_{max}+i,k]$ ensures that the final point(s) of the backwards summation are still incorporated into the computation even if the arbitrarily chosen value of $a$ does not generate a final increment length that ends on $k$. $\square$

Note that $D$ is not explicitly needed for computing equation \ref{eq:adapttime} because the distances $d$ are implicitly taken into account by $\zeta$. Using the median value of each increment avoids the need for interpolation between time points. The implementation of the adaptive memory method described here is necessarily limited to smooth functions due to the assumption that the neighbors of the median values used in the summation do not vary much over the increment being considered (but see Discussion below). This method essentially allows for a contribution by all previous time points to the current calculation, yet reduces computational times by minimizing the total number of calculations. For the interested reader the full source code for the implementation is available online from the authors' website at www.silva.ucsd.edu.

\begin{figure}
\begin{center}
\includegraphics[width=8.8cm]{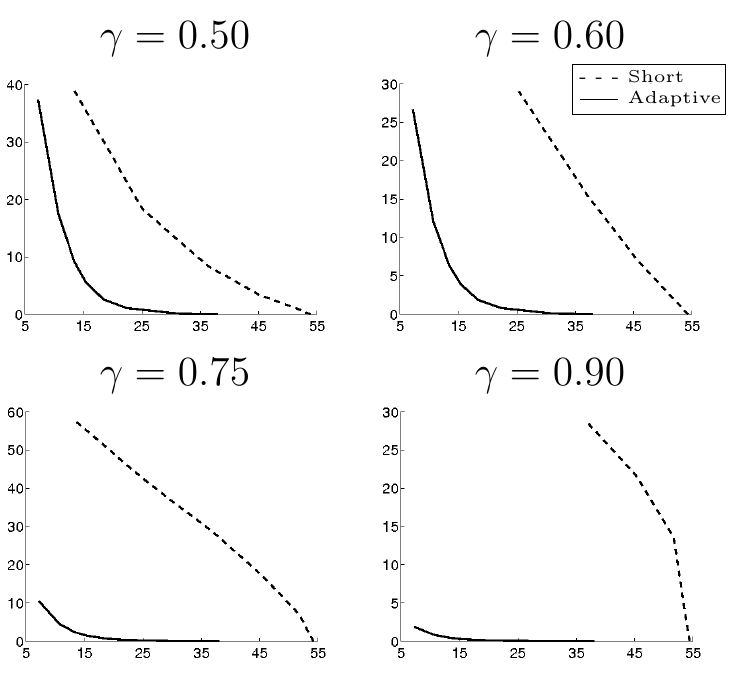}    
\caption{Comparison of the error between adaptive memory and the short memory as a function of the calculation time (x-axis: computation run time in seconds) expressed as a percentage error relative to the computed value for the full non-shortened approach (y-axis). Four different values of $\gamma$ are shown.} 
\label{shortvsadaptive}
\end{center}
\end{figure}

The results of using various $L$ (for short memory) and interval steps (for adaptive memory) are shown in Fig.~\ref{shortvsadaptive}. Increasing the values of $L$ and $a$ resulted in a decrease in the error of the estimated results but at the cost of increased computation times.  Conversely, decreasing the values of $L$ and $a$ resulted in a decrease in computation times, but at the expense of accuracy.  In all cases however, the adaptive time method had a significantly lower error for the same simulation time, and also reached a clear asymptotic minimum error much faster then the minimum error achieved by the short memory method. In these simulations, $\alpha=1$, $\beta=0$, $\Delta_t=1$, $\Delta_x=10$, using a 20 x 20 grid, and ran until $t=1500$ where $U_{10,10}^0=10$. 

The error for the `short memory' method increased comparatively quickly and got worse as $\gamma\rightarrow1$. This was due to the fact that the diffusion from the initial point source was initially quite fast near $t=0$, which were the first time points cut by the `short memory' algorithm. Because the component near $t=0$ represents the majority of the immediate past history function, large errors result even when $L\rightarrow t$.

\section{Discussion}

The regular diffusion equation has been a cornerstone of transport modeling for decades. Regular diffusion relies on the assumption of a Markovian process that is not necessarily present in natural systems.

One approach to modeling these non-markovian processes is using the fractional diffusion equation (eq. \ref{fracdiff}). Mathematically such methods have been around for decades but it is relatively recently that they have been applied to the natural sciences. Computationally efficient methods for numerically evaluating these equations are a necessity if fractional calculus models are to be applied to modeling real world physical and biological processes.

It should be noted that while in this work the simulations were done in the subdiffusive regime for a simple point source, the methods we derive here are directly applicable to more complex sources or superdiffusion ($\gamma>1$). However, complex fast-evolving past histories in the form of a forcing function $(f(u))$ or oscillations generated in a superdiffusion regime will result in much larger errors for both the short and adaptive memory methods. In the case of the adaptive memory method introduced in this paper this is due to its `open-loop'-like algorithm that blindly increases the spacing between points in the summation as the calculation goes further back in time and which does not take into account the speed of evolution of the equation. Adaptive time approaches for regular differential equations often make the integration step size a function of the derivative, i.e. more closely spaced time steps are used when the function is oscillating quickly and cannot be averaged without loss of accuracy, and widely spaced time steps are used where the function is smooth. In the current implementation we have assumed that the past history function $\psi(\gamma,m)\delta^{k-m}_{i,j}$ is smooth. Subsequent work will implement a `smart' `closed-loop'-like algorithm where the step size of the backwards summation is dependent on the derivative of the past history function, i.e. a form of feedback. This would optimize the speed of the simulation while reducing the error due to the averaging of time points in the backwards summation, ultimately resulting in low errors for both high frequency forcing functions in the subdiffusion regime and for oscillations intrinsic to the superdiffusion regime.

\begin{ack}
We would like to thank Dr. Richard Magin and Dr. Igor Podlubny for helpful discussions during the course of this work.
\end{ack}

\bibliography{fracdiffbib}             

                                                                        
\end{document}